%% file: main.tex
\documentclass[10pt,conference]{IEEEtran}
\include{include}

 \hypersetup{ 
    colorlinks,% 
    citecolor=black,% 
    filecolor=black,% 
    linkcolor=black,% 
    urlcolor=black,%
pdftitle={Social Event Scheduling},
pdfauthor={Nikos Bikakis, Vana Kalogeraki, Dimitrios Gunupulos},
pdfsubject={Social Event Planning Arrangement Organization  Profit, Attendance Maximization},
pdfproducer={pdfbikTeX-1.1},
pdfcreator={bikTeX},
pdfkeywords={Social Event Planning, Social Event Arrangement, Social Event Organization  SES problem, Social Event Mining, 
profit maximization, attendance maximization, user preferences, meetup, Event-based Social Networks, EBSN, event recommendation, social networks,
assignment problem, matching problem, event organizers, event planning, marketing, event scheduling, nphard, knapsack, np-hard,
approximation, optimization, large data, big data, data science, data mining, greedy, user habits, user behavior, spatiotemporal, 
given a set of events a set of time periods and a set of users
our objective is to determine how to assign events on the time periods
so that the maximum participant enrollment is achieved, 
the Social Event Scheduling SES problem,  
which schedules a set of social events considering 
user preferences and behavior, events spatiotemporal conflicts, and competing events, 
in order to maximize the overall number of participants.
}
}

\IEEEoverridecommandlockouts

\begin{document}

\title{Social Event Scheduling {\Huge$^\star$}\vspace*{-0pt}\thanks{$^\star$
\textbf{This paper appears in 34th IEEE Intl.\ Conference on  Data Engineering (ICDE 2018)}\vspace*{-19pt}}
}

% 
%\author{\IEEEauthorblockN{Nikos Bikakis Vana Kalogeraki}
%\IEEEauthorblockA{Athens University of  Economics \& Business,  Greece\\
%Greece\\
%line 4: Email: name@xyz.com}
%\and
%\IEEEauthorblockN{Dimitrios Gunopulos}
%\IEEEauthorblockA{University of  Athens, Greece\\
%Greece\\
%line 4: Email: name@xyz.com}
%}

\author{%
% author names are typeset in 11pt, which is the default size in the author block
{Nikos Bikakis{\small $~^{\dagger}$}, 
Vana Kalogeraki{\small $~^{\dagger}$}, 
Dimitrios Gunopulos{\small $~^{\#}$} }%
\vspace{1.6mm}\\
\fontsize{10}{10}\selectfont\itshape
$^{\dagger}$Athens University of  Economics \& Business,  Greece
%$^{\star}$\,Athens University of Economics \& Business, Greece
\hspace*{5pt}
$^{\#}$University of  Athens, Greece\\
\fontsize{9}{9}\selectfont\ttfamily\upshape
bikakis@dblab.ntua.gr \hspace*{2pt}
   vana@aueb.gr  \hspace*{2pt}
   dg@di.uoa.gr 
  %$^{2}$\,second.author@second.com
}

\newcommand{\alt}[2]{% {short}{full}
#1%% short version
%#2%% full version
}

\maketitle

%\vspace*{-10pt}

\begin{abstract}
%Event-based Social Networks (EBSN) have become extremely popular in recent years. 
%In this new type of social media, users organize, manage and share social events. 
%In conjunction with EBSNs, several
% entities such as event planning and marketing companies, organizations, as well as venues, organize and manage numerous social events (e.g., festivals, conferences, parties). 
 
A major challenge for social event organizers (e.g., event planning and marketing companies, venues) is attracting the maximum number of participants, 
since it has great impact on the success of the event, 
and, consequently,  the expected gains (e.g., revenue, artist/brand publicity).
In this paper, we introduce the   \textit{Social Event Scheduling} (\ses) problem,  
which schedules a set of social events considering 
user preferences and behavior, events' spatiotemporal conflicts, and competing events, 
in order to maximize the overall number of attendees.
We show that \ses is strongly NP-hard, even in highly restricted instances.
 To cope with the hardness of the SES problem we design a greedy approximation algorithm. 
Finally, we evaluate our method experimentally using a  dataset from the Meetup event-based social network.%

 \end{abstract}


%
\input{intro}

\input{problem}

\input{methods}

\input{eval}

%\input{related}
\input{concl}

 \section*{Acknowledgment}
This research has been financed by the European Union through the FP7 ERC IDEAS 308019 NGHCS project, the Horizon2020 688380 VaVeL project and a 2017 Google Faculty Research Award.

\vspace{-3pt}

%\bibliographystyle{abbrv}
%\bibliography{biblio}

\input{bibshort.bbl}
%\footnotesize
%\small
%\bibliographystyle{IEEEtran}
%\bibliographystyle{abbrv}
%\bibliography{biblio}

%\alt{
%\input{bibshort.bbl}
%}
%{
%\bibliographystyle{abbrv}
%\bibliography{biblio}
%}

% \newpage
\normalsize
\alt{
%\appendix
%\input{appendix-short}
}{
\appendix
\input{appendix}
 }

\end{document}

%% file: include.tex
\usepackage[makeroom]{cancel}
\usepackage{pifont}
\usepackage{marvosym}

 \usepackage[justification=centering]{caption} 
\usepackage[table]{xcolor}
 \usepackage[caption=false]{subfig}
\usepackage{color}
\usepackage{multirow}
 \usepackage{graphicx}
\usepackage[linesnumbered, vlined, ruled,algo2e]{algorithm2e} 
\usepackage{booktabs}
\usepackage{amssymb}
\usepackage{amsfonts}
\usepackage{amsmath}
\usepackage{mathtools}

\usepackage{amsthm}
\usepackage{soul}
\usepackage{etex}
 \usepackage{relsize}
\usepackage{epstopdf}
\usepackage{multirow}
\usepackage{url}
\usepackage{tablefootnote}
\usepackage{setspace}
\usepackage[bookmarks=false]{hyperref}
\hypersetup{ 
    colorlinks,% 
    citecolor=black,% 
    filecolor=black,% 
    linkcolor=black,% 
    urlcolor=black,%
}

 \usepackage{cite}

\newcommand{\eat}[1]{}

\newcommand{\stitle}[1]{\vspace{0.2cm}\noindent\textbf{#1}} %5.5 pt

\DeclareMathOperator*{\argmax}{arg\,max}

\renewcommand{\qed} {\hfill$\blacksquare$}

\definecolor{dark-gray}{gray}{0.2}

\newcommand{\ses}{{SES}\xspace}
\newcommand{\bsc}{{GRD}\xspace}

\newcommand{\rand}{{RAND}\xspace}
\newcommand{\stat}{{TOP}\xspace}

\newcommand{\popAss}{{\mathsf{popTopAssgn}}}

\newcommand{\E}{\ensuremath\mathcal{E}}

\renewcommand{\S}{\ensuremath\mathcal{S}}

 \renewcommand{\L}{\ensuremath\mathcal{L}}
 \newcommand{\C}{\ensuremath\mathcal{C}}
 \newcommand{\T}{\ensuremath\mathcal{T}}
 \newcommand{\U}{\ensuremath\mathcal{U}}

\newcounter{prob}
\newenvironment{myprob}[1][]
{\refstepcounter{prob}\par\setlength{\leftskip}{6pt}\setlength{\rightskip}{0pt}\medskip\noindent\ignorespaces
\textbf{#1}}
{\medskip\par}

\newcounter{theorcnt}
\newenvironment{mytheor}
{\refstepcounter{theorcnt}\medskip\setlength{\leftskip}{6pt}\setlength{\rightskip}{0pt}\par\noindent\ignorespaces 
   \textbf{Theorem~\thetheorcnt.}}
{\smallskip\par}

\newenvironment{myproofSk}
{\medskip\setlength{\leftskip}{0pt}\setlength{\rightskip}{0pt}\par\noindent\ignorespaces 
   \textsc{Proof Sketch.}}
{\medskip\par}

\newcounter{corollcnt}

\newcounter{claimcnt}

\newcounter{remarkcnt}
\newenvironment{myremark}
{\refstepcounter{remarkcnt}\vspace{0.2cm}\setlength{\leftskip}{0pt}\setlength{\rightskip}{0pt}\par\noindent\ignorespaces 
   \textbf{Remark~\theremarkcnt.}}
{\smallskip\par}

\newcounter{proposcnt}

\newcounter{lemmcnt}
{\medskip\par}

\newcounter{ex}

\newcommand{\superscript}[1]{\ensuremath{^{\textrm{#1}}}}

\newcommand{\myFontP}[1]{{\fontfamily{ppl}\selectfont #1}} %Adobe Palantino 
\newcommand{\tline}  {\specialrule{0.8 pt}{0pt}{1pt}}		 %top table line 
\newcommand{\bline}  {\specialrule{0.8 pt}{1pt}{0pt}}		%buttom table line 
\newcommand{\dline}  {\specialrule{0.4 pt}{0pt}{2pt} \specialrule{0.4 pt}{0pt}{4pt}}		%double table line 
\newcolumntype{'}{!{\vrule width 0.6pt}}

\SetKw{Logand}{and}	
\SetKw{Logor}{or}
\SetKw{Lognot}{not}	

\SetKw{mybreak}{break}	%define your own keywords
\SetKw{mycontinue}{continue}	%define your own keywords

\SetKwInput{KwVar}{Variables}
\SetKwInput{KwPar}{Parameters}

\SetKwData{myEOF}{EOF}	
\SetKwData{myfalse}{false}
\SetKwData{mytrue}{true}
\SetKwData{mynull}{null}
\SetKwData{myvalid}{\texttt{valid}}
\SetKwData{myfeasible}{\texttt{feasible}}
\SetKwData{myupdated}{\texttt{updated}}
\SetKwData{mynotupdated}{\scalebox{0.9}[1]{ \texttt{prtl  \!\! \!\!\!updated}}}

\SetKwData{myempty}{empty}
\SetKwBlock{myblock}{beginnikos}{end} %define your own keywoed blocks 
\SetAlgoInsideSkip{} 
\SetAlgoSkip{} 
\SetVlineSkip {0.5mm} 
\SetAlgoCaptionSeparator{.} 

\newcommand{\mycomment} [1] 	{\tiny{\textcolor{dark-gray}{\myFontP{{#1}}}}} %\textcolor{dark-gray}
\SetKwComment{Comment}{\mycomment{/\!\!/}}{}
 \DontPrintSemicolon
\SetAlgoCaptionLayout{small}
\SetProcNameSty{small}
\SetProcArgSty{small}
\graphicspath{{figures/}{plots/}}

%% file: intro.tex
%!TEX root = main.tex 

\vspace*{-13pt}
\section{Introduction}
\label{sec:intro}

 The wide adoption of social media
%The proliferation of social media
 and networks has recently given rise 
to a new type of social networks that focus on online event management,
called  Event-based Social Networks (EBSN) \cite{Liu2012}. 
In the most predominant EBSN platforms, such as 
\alt{Meetup,  Eventbrite or Whova,}
{Meetup\footnote{\small\href{ https://www.meetup.com}{www.meetup.com}},  
Eventbrite\footnote{\small\href{https://www.eventbrite.com}{www.eventbrite.com}}
or Whova\footnote{\small\href{https://whova.com}{http://whova.com}},}
users  {organize},  {manage} and  {share} social events and activities. 
In conjunction with the events' organizers in \mbox{EBSNs}, several entities 
such as  \textit{event planning} and \textit{marketing companies} 
\alt{
(e.g., Jack Morton, GPJ),}
{
(e.g., Jack Morton\footnote{\small\href{www.jackmorton.com}{www.jackmorton.com}}, GPJ\footnote{\small\href{www.gpj.com}{www.gpj.com}}),}  
\textit{organizations} (e.g., IEEE), 
as well as  {\textit{venues}} (e.g., theaters, night clubs), organize and manage a  {variety} of social events (music concerts, conferences, promotion parties). 
%A major challenge for event organizers is attracting the maximum number of participants, 
%as it has great impact on the success of the event, 
%and, consequently, on the profits earned from the event.
\textit{A major challenge for event organizers is attracting the maximum number of participants,  since it has great impact on the success of the event, 
and consequently, on  the expected gains from it, for all involved
(e.g., revenue, artist/brand publicity)}.

%\RED{SAY AMAZON + FCB}

% 
%\RED{AMAZON IS GETTING INVOLVED IN THE MUSIC FESTIVAL BUSINESS
% offering ON-SITE services 
%improve” the experience of the 32 million people that attend music festivals in the US each yea}

Consider the following real-world scenario. 
%\alt{
%A company is going to organize  the \textit{Summerfest} festival.}
{A company is going to organize  the \textit{Summerfest} festival.
%\footnote{ \href{http://summerfest.com}{http://summerfest.com}}.
%}
%a festival to take place during the first week of August.  
%Every day of said week several multi-themed social events will be taking place.  
%It is worth to mention an example of such a festival.
Summerfest is an 11-day music festival, featuring   11 stages and attracting more than 800K people each year. 
Throughout the festival, in addition to the music concerts, numerous multi-themed events take place (e.g., theatrical performances). 
%ranging from art-makings and theatrical performances to children games and fitness activities.art-makings, 
%
Assume that Alice enjoys listening to Pop music, and is a fashion lover. 
On Monday from 7:00 to 10:00pm, a concert of a famous Pop band is scheduled to take place at the festival. 
At the same day, on a different stage, a fashion show is taking place from 7:00 to 9:00pm. 
Furthermore, from 6:00 to 8:00pm on that day, a music concert of a Pop singer has been organized by a nearby (competing) venue.
Despite the fact that Alice is interested in all three events, she is only able to attend one of them. 
In another scenario, assume that a Pop concert is hosted by the festival on Tuesday evening, 
but Alice is not capable of attending this event, because on Tuesdays she works until late at night.

The above example illustrates the major aspects that should be considered 
in events scheduling scenarios.
In order to attract as many attendances as possible, 
organizers have to carefully select the events that are going to take place during the festival, 
possibly picking among from numerous candidate events, as well as the date/time on which 
each event is going to take place. 
During the event  scheduling process, at least the following aspects have to be 
considered: \textit{user preferences}, \textit{user {habits}} (e.g., availability), 
 \textit{spatiotemporal  conflicts between scheduled events}, and \textit{possible third parties  events} (e.g., organized by a third party   company) which might attract potential attendees (i.e., competing events).

In this work, we introduce the \textit{Social Event Scheduling} (\ses) 
problem, which considers the aforementioned aspects 
and the goal is to maximize the overall number of participants in the scheduled events.
\textit{In short, given a set of events, a set of time periods and a set of users, 
our objective is to determine how to assign events on the time periods, 
so that the maximum participant enrollment is achieved}.
%
 
%\RED{ In what follows, we present a simple example that introduces 
% the main entities involved in \ses problem.} 

Recently, a number of works have been proposed in the context of  event-participant planing \cite{Li2014,She2015a,She2016,She2015,Tong2015,She2017,Huang08092016,ChengYCGW17}. %  
These works examine a problem from a different perspective:
given a set of {pre-scheduled} events, they focus on finding the most appropriate
assignments for the users (i.e., participants) attending the events.
The determined \textit{user-event assignments aim at maximizing the satisfaction of the  users}.
However, these works fail to consider a crucial issue in event
management, which is the ``satisfaction'' (e.g., revenue, publicity)
of the entities involved in the event    organization (e.g., organizer, artist,  sponsors, services' providers). 
Here, in contrast to existing works, \textit{our objective is to maximize the
satisfaction of the event-side entities.}
To this end, instead of assigning users to events, \textit{we assign events to time
intervals, so that the number of events’ attendees is maximized}.
Briefly,  we study  an ``\textit{event-centric}"  problem, while the existing approaches focus on ``\textit{user-centric}" problems.

Therefore, our \textit{objective} is substantially different compared to the existing works.
The same holds for the \textit{solution}; in our problem, the {solution} is a set of  \textit{event-time assignments}, while in existing works is a set of \textit{user-event assignments}.
%
%Some further differences with existing \textit{problems' settings} are the following.
% 
Additionally, in order to solve our problem we have to find a subset from a set of candidate events (i.e., some events  may not be included in the solution), while in other works the solution contains all the users (i.e., each user is assigned to events).
Finally, beyond the user and event entities which are also considered in   existing works, in our problem more core entities are involved (e.g., event organizer, competing events).
Thus, overall, the \textit{objective}, the  \textit{solution} and   the \textit{setting} of our problem 
substantially differ from existing works.

\alt{}{Furthermore, the problem studied here has common characteristics with the {Generalized assignment} (GAP) and {Multiple knapsack} (MKP) problems \cite{Martello1990}. 
However, a major difference of \ses compared to GAP (and MKP) is that in \ses the 
the utility (i.e., expected attendance)
of assigning an event (resp.\ item) to an interval (resp.\ bin) is defined w.r.t.\ the other events assigned to this interval (see Section~\ref{sec:anal} for more details).}
	
%\RED{
% The main contributions of this work are summarized as follows:
%%
%$(1)$~We introduce the  \textit{Social Event Scheduling} (\ses) problem;
%%, which finds a schedule for a set of events, so as the overall  number attendances is maximized;
%%
%$(2)$~We show that the \ses problem is strongly \mbox{NP-hard}, a well as \mbox{NP-hard} to be approximated over a factor;   % larger than $(1- \epsilon)$. 
%%
%  $(3)$~We design four approximation algorithms. 
%These algorithms  exploit a series of  schemes 
%that we develop to improve their performance.}
%% 

%% file: problem.tex
%!TEX root = main.tex 

\alt{}{
\begin{table}[]
\centering
\caption{Common  notation}
\label{tab:notation}
%\footnotesize
\scriptsize
% \small
\begin{tabular}{cl}
\tline
\textbf{Symbol} & \textbf{Description}\\ \dline
$\theta$ & Number of available resources \\ \rowcolor{gray!10}	
$\T, t$  & Set of time intervals,  a time interval\\
$\E, e$  & Set of candidate events, a candidate event\\ \rowcolor{gray!10}	
$\ell_e$ & Location of  $e$\\
$\xi_e$ & Number of resources required for $e$\\ \rowcolor{gray!10}	
$\C, c$ & Set of competing events, a competing event \\
$t_c$ & Time interval associated with competing event $c$\\ \rowcolor{gray!10}	%remove 
$C_t$ & Set of  competing events associated with  $t$\\ %remove 
$\S$  & Event scheduling \\\rowcolor{gray!10}	
$\alpha_e^t$ & Assignment of $e$ at $t$\\
$t_e(\S)$ &   The time interval that $\S$ assigns to $e$\\ \rowcolor{gray!10}	 %remove 
$E(\S)$ & Set of events that $\S$ assigns time intervals \\ %remove 
$E_t(\S)$ & Set of events that $\S$ assigns to $t$ \\ \rowcolor{gray!10}	
$T(\S)$ &  Set of time intervals that $\S$  have assigned events \\ %remove 
$\U, u$ & Set of users, a user \\ \rowcolor{gray!10}	
$\mu_{u, h}$ & Interest of   $u$ over (candidate or competing) event $h$\\
$\sigma_u^t$  & Probability of $u$ participating in a social activity at $t$\\  \rowcolor{gray!10}	
$\rho_{u, e}^t$  &  Probability of $u$   attending $e$   at $t$ \\
$\omega_{e}^t$  & Expected attendance of $e$ at $t$ \\  \rowcolor{gray!10}	
 
 $g_{e}^t$  & Score (i.e., utility gain) of assignment  $\alpha_e^t$ \\ 

$\Omega({\S})$ & Total utility  for $\S$ \\  \rowcolor{gray!10}	
$\Phi$ & Score bound  \\ %remove
 \bline
\end{tabular}
 \end{table}
}

 \section{Social Event Scheduling Problem}
\label{sec:prob}

In this section we first introduce the \textit{Social Event Scheduling} 
(\ses) problem; and then we study its complexity. 
Before we formally introduce our problem, we present some
necessary definitions.
\alt{}{Table~\ref{tab:notation} summarizes the most common notation and their
description.}

 \stitle {Organizer \& Time intervals.}
 We assume that the event \textit{organizer} (e.g., company, venue) 
 is associated with a number of  (available) {\textit{resources}} $\theta \in \mathbb{R}^+$. 
For example, as resources we can consider the agents (i.e., staff) 
which are responsible to setup and coordinate the events.
 % is the team formed by $\theta \in \mathbb{N}^+$ \textit{agents}. 
Let $\T$ be a set of \textit{candidate time intervals}, 
representing time periods that are available for organizing events.
%In each time interval a user can attend at most one event.
 %
Note that  the intervals contained in $\T$ are disjoint.

%However, intervals do not strictly define the beginning and the end of the events.
%Thus, overlapping events can take place in the same interval.   
%Further, it is apparent that in each time interval the user can attend at most one event. 

 \stitle {Candidate  Events.}
Assume  a set  $\E$  of available events to be scheduled, referred as \textit{candidate  events}.
Each $e \in \E$ is associated with a \textit{location} $\ell_e$ 
representing the place (e.g., a stage) that is going to host the event.  
%Also, $e$ is associated with a number $\xi_e \in \mathbb{R}^+_0$, 
%defining the number of \textit{resources required} for  $e$.
Further, each event $e$ requires a specific amount of resources $\xi_e \in \mathbb{R}^+_0$  for its  organization, referred as \textit{required resources}.

% Note that more than one locations (e.g., rooms) may associated with a single venue (e.g., cinema). 
% Also, $e$ is associated with a set of time intervals $A_e \subset \T$, referred
%  as \textit{availability} set, defining the intervals in which $e$ can take place.
% For example, considering a music concert, 
% the band that will perform might be available only at specific periods, or 
% there are periods in which the concert's venue is reserved for other events. 

\stitle {Schedule \& Assignment.} 
An \textit{assignment} $\alpha_e^t$ denotes that the candidate event $e \in \E$ is scheduled to take place at $t \in \T$.
An event \textit{schedule} ${\S}$ is a set of assignments, where there exist no two assignments referring to the same event. 
Given a schedule $\S$,  we denote as $\E(\S)$ the set of all candidate events that  are scheduled by ${\S}$; 
 and $\E_t(\S)$ the candidate events that  are scheduled  by ${\S}$  to take place at $t$ (i.e., assigned to $t$).
 Formally,  $\E(\S)=\{ e_i \in \E \mid a_{e_i}^{t _i }\in \S\}$ and 
$\E_t(\S)=\{ e_i \in \E \mid a_{e_i}^{t _i }\in \S \text{ with }  t_i=t \}$.
Further, for a candidate event $e \in \E(\S)$, we denote as
 $t_e(\S)$ the time interval on which $\S$ assigns $e$.

%
%An event \textit{schedule} ${\S}\colon \E \to \T$,  is a partial  function which assigns 
%%Vana: to egrapsa event schedule anti gia event scheduling
%a candidate event $e \in \E$ to a time interval $t \in \T$.
%An \textit{assignment} $\alpha_e^t$, means that the candidate event $e$ is scheduled to take place at $t$.
%%
%For  a schedule $\S$, we denote
%as $t_e(\S)$ the time interval that $\S$   assigns the candidate event $e$; 
%while for an  event $e'$ that is not assigned to any time interval, we have that ${t_{e'}(\S)=\varnothing}$.
%%
%Further, let $\E(\S)$ denote the candidate events that ${\S}$  assigns to a time interval, 
% and $\E_t(\S)$ the candidate events that  ${\S}$ assigns to the time interval $t$.
%%and $T({\S})$ the time intervals from $\T$ that events have been assigned.
%Formally, 
%${\E(\S)=\{e \in \E \mid t_e(\S) \neq \varnothing \}}$ and
%$\E_t(\S)={\{e \in \E \mid t_e(\S) = t\}}$.
%%and  ${T({\S})=\{t \in \T \mid E_t(\S) \neq\varnothing \}}$.

 \stitle {Feasibility.}
A schedule $\S$ is said to be \textit{feasible} 
if the following constraints are  satisfied: 
% (1) $\forall  t \in R(\S)$ and  $\forall e \in t(\S)$   holds that $e(\S) \in A_e$ (\textit{availability constraint});
(1)~$\forall  t \in \T$ holds that 
$\nexists  e_i, e_j \in \E_t(\S)$ with  $\ell_{e_i} = \ell_{e_j}$ 
% and  $\forall e_i, e_j \in E_t(\S)$ holds that $\ell_{e_i} \neq \ell_{e_j}$ 
(\textit{location constraint}); and 
(2)~$\forall t \in \T$ holds that 
${\underset{\forall e \in \E_t(\S)}{ \!\! \!\!\!\! \!\! \sum} \!\!\!\!\!\!\!\! \xi_e} \leq \theta$
%$ \sum_{\forall e \in E_t(\S)}  \xi_e \leq \theta$
%(\textit{agent availability  constraint}).
(\textit{resources constraint}).
%Note that we assume that there exists at least one feasible scheduling $\S$ so that  $R(\S) \neq \varnothing$.
In analogy, an \textit{assignment} $\alpha_e^t$ is said to be \textit{feasible} if the aforementioned constraints hold for $t$.
Further, we  call \textit{valid assignment}, 
an assignment $\alpha_e^t$ 
when the assignment is \textit{feasible} and $e \notin \E(\S)$.

 \stitle {Competing Events.}
 Let $\C$ be a set of \textit{competing events}. 
As competing events we define events that have already been scheduled by third parties (e.g., organized by a third party marketing company), and will possibly attract potential attendees of the candidate events.  
Based on its scheduled time, each competing event 
$c \in \C$ is associated with a time interval $t_c \in \T$. 
\alt{% This association implies a ``conflict" between the competing event $c$ and 
%and a possible candidate event that will take place at the time interval $t_c$. 
%Note that,  at each time interval, a user can attend at most one event. 
}
{This association implies a ``conflict" between the competing event $c$ and 
and a possible candidate event that will take place at the time interval $t_c$. 
Note that,  at each time interval, a user can attend at most one event. }
Further, as $\C_{t}$ we denote  the competing events that are associated with 
the time interval $t$; i.e., ${\C_{t}=\{c \in \C \mid t_c=t\}}$.

 \stitle{Users.}
Consider a set of {users} $\U$, for each \textit{user} $u \in \U$ and event $h \in \E \cup \C$, 
there is a function ${\mu \colon \U \times (\E \cup \C) \to [0, 1]}$, 
denoted as  $\mu_{u,h}$,  that models the \textit{interest} of user  $u$ over $h$. 
The interest value (i.e., affinity)  can be estimated by considering a large number of  factors 
(e.g.,   preferences, social connections)\superscript{\ref{fot:mining}}.  %, spatial information
%\cite{She2015a,ZhangWF13,Zhang2015}.
%However, the calculation of this value is beyond the scope of this work.
%(e.g., see \cite{She2015a,ZhangWF13,Zhang2015}). %Tong2015 DuYMWWG14
%

Moreover,  for each user $u$ and time interval $t$ 
a \textit{social activity probability} $\sigma_u^t$ is considered, representing the {probability of user} $u$ {participating in a social activity at} $t$. 
Formally we have ${\sigma \colon \U \times \T \to [0, 1]}$.
 This probability   can be estimated by examining the user's past behavior (e.g., number of check-ins)\superscript{\ref{fot:mining}}.
% \cite{ZhangWF13,Boutsis15,Zhang2015}. %DuYMWWG14
 %(e.g., large number of check-ins on Mondays).
Note that, user data  can either be gathered by analyzing organizer data 
(e.g., registered users profiles) or be provided by a market research company.

  \stitle {Attendance.}
Assume a user $u$  and a candidate event $e \in \E$ 
that is scheduled by $\S$ to take place at  time interval $t$; 
$\rho_{u,e}^t$ denotes the  \textit{probability of} $u$  \textit{attending} $e$  \textit{at} $t$.
Considering the \textit{Luce's choice theory}, %\cite{Luce1959}, 
%Recall that, at each time interval, a user can attend at most one event. Hence,
 the probability $\rho_{u,e}^t$ is influenced  by  the
social activity probability $\sigma$ of $u$ at $t$, and
the interest $\mu$ of $u$ over $e$, $\C_t$ and $\E_t(\S)$.
%  and the interest of $u$ over $C_t$ and $E_t(\S) \backslash e$.
%
We define the \textit{probability of} $u$  \textit{attending} $e$  {at} $t$  as%
{\footnote{Event-based mining methods 
can be used to compute this value, e.g., 
%\cite{Zhang2015}. 
\cite{ZhangWF13,Boutsis15,Zhang2015,Romero2017,FengCBM14,XuZZXCL15,DuYMWWG14}.
However, this is beyond the scope of this work. \label{fot:mining}
}:

{ 
%\vspace{-3pt}
\small{
 \begin{equation}
% \begin{center}
\rho_{u,e}^t =  \sigma_u^t \:  \dfrac{\mu_{u,e}}{
{\underset{\forall c \in \C_t}{ \sum}\mu_{u,c}} +
{\underset{\forall p \in \E_t(\S)}{ \sum} \! \!  \! \! \mu_{u,p}}} 
 %\end{center}
 \label{eq:att}
  \end{equation}
}
%\vspace{-2pt}
}

\noindent
 Furthermore, considering all users $\U$,  we define the 
\textit{expected  attendance}    for an event  $e$ scheduled to take place at   $t$ as:

%\vspace*{-4pt}
{ \small{
 \begin{equation}
     \omega^t_{e} =  \underset{\substack{\forall u \in  \U}}{ \sum} \:\:  \rho_{u,e}^t 
\label{eq:user}
  \end{equation}
}}

 \stitle {Total Utility.}
The \textit{total utility} for a schedule $\S$, denoted as $\Omega(\S)$,
is computed by considering the expected attendance over all scheduled events.
Thus, we have: 

%\vspace{-4pt}
{ \small{
 \begin{equation}
%\begin{center}
     \Omega(\S) =  
    \underset{\substack{ \forall e \in  \E(\S) }}{ \sum} \:
    \omega^{t_e(\S)}_{e}  
\label{eq:total}
%  \end{center}  
 \end{equation}
}}

%  \vspace{-3pt}

\noindent
We formally define the \textit{Social Event Scheduling} (\ses)  problem as follows:
\vspace{-3pt}
\begin{myprob}[Social Event Scheduling  Problem (\ses).]
Given an \textit{integer} $k$, a set of \textit{candidate time intervals} $\T$; 
a set of \textit{competing events} $\C$;
% where each $c \in \C$ is associated with a  time interval $t \in \T$;
a set of \textit{candidate events} $\E$; and a set of  \textit{users} $\U$;
%and their associated attributes;
our goal is to find a \textit{feasible schedule} $\S_k$ that determines how to assign $k$ candidate events such that the \textit{total utility} $\Omega$ is maximized; i.e.,  ${\S_k = \argmax \Omega({\S})}$ and  
%$\mid E(\S) \mid =\Min(k, |\E|)$.
$|\S|=k$.  
\end{myprob}

\eat{
\stitle{Remark.}
A simple ``profit-oriented" version of the \ses problem may also be defined
by associating each event with   an organization cost and a fee.
These factors can be directly included in the \ses problem.
In this setting, the total utility will determine the  expected scheduling profit. 
Also, several additional  constraints  may be included, 
e.g., associate  events with duration, or  
a set of time intervals during which 
the event's location can host the event, etc.
}

%to calculate the expected attendance
% capacity 
% duration
% 

% Given a number of agents $\theta$, 
% candidate time intervals $\T$, 
% candidate events $\E$, where each $e \in \E$ has a locations $L_e$,   and a number of required agents $\xi_e$; 
% competing events $\C$, where each $c \in \C$ is associated with a  time interval $t \in \T$; 
% and users $\U$ where each user has an interest function $\mu$, as well as a social activity probability $\sigma$; 
% find a feasible scheduling $\S^*$ that maximizes the total utility  $\Omega({\S})$; i.e., 
% ${\S^* = \argmax \Omega({\S})}$.

% Note that a constraint-free version of the \ses problem can be defined by letting $A_e=\varnothing$, $L_e=\varnothing$, $\xi_e=0$, $\forall e \in \E$; and $\theta=\alpha$, where $\alpha>0$.
%

\alt{}
{Without loss of generality, in  the \ses definition we have assumed that $|\E| \ge k$ 
and there exists at least one feasible assignment for $k$  events. 
}

Next,  we show that even in highly restricted instances the \ses problem is \textit{strongly NP-hard}.

%\noindent
%First, we show that the  \ses problem is strongly NP-hard even in highly restricted instances%
%\footnote{ Due to lack of space,   we only include proof sketches, 
%while in simple cases, the proof sketch is also omitted.}.

\begin{mytheor}
\label{th:nphard}
The \ses problem is strongly \mbox{NP-hard}.
\end{mytheor}

\vspace{-2pt}
\begin{myproofSk}
Our reduction is from  the Multiple Knapsack Problem with Identical bin capacities (MKPI),
which is known to be strongly \mbox{NP-hard} \cite{Martello1990}.
In MKPI there are multiple items and bins. 
Each item has a  weight and a profit  and all bins have the same capacity. 
In the reduction, we use the  following \textit{associations  between the} MKPI \textit{and the} \ses:
$(1)$~bins to time intervals;
$(2)$~capacity to number of available resources;
$(3)$~items to events; 
$(4)$~weight  to number of required resources;  
$(5)$~item profit  to likeness; and
$(6)$~total profit to expected attendance.
Further, in the proof, we consider the following \textit{restricted instance of} \ses:
$(1)$~the users are as many as the candidate events;
$(2)$~there is only one competing event in each time interval; 
$(3)$~all users have the same interest value $K$ over the competing events;
$(4)$~each user likes only one event and  each event is liked only by one user; 
$(5)$~the interest function is ${\mu= p\frac{K}{1-p}}$,  where   $p$ is item profit;
$(6)$~the social activity probability  is the same for each user and time interval; and
$(7)$~there are no  location constraints.
 \qed
\end{myproofSk}
%
% $(7)$~The interest function $\mu$ is specified as follows.  
%
% $(7a)$~Regarding competing events, for each $c \in \C$ and $\forall u \in \U$ holds that $\mu_{u,c} = K$. 
%
% $(7b)$~Further, regarding candidate events, we have that each user likes only 
%one candidate event and each candidate event is liked  only by one user. 
%Let a user $u$ that expresses her interest over the event $e_u$.
%Then the interest value $\mu_{u, e_u}$ is defined using 
%profit   $p(i_u)$,  where $i_u$ is the item that corresponds to the event $e_u$. 
%Particularly,  we have that ${\mu_{u, e_u}= p(i_u)\frac{K}{1-p(i_u)}}$, 
%and ${\forall e' \in \{\E  \backslash e_u\}}$,  $\mu_{u, e'}= 0$. 

%% file: methods.tex
%!TEX root = main.tex 

 \section{Greedy Algorithm (\bsc)}
\label{sec:methods}
%\vspace{-5pt}
 
%

First, we define the assignment score; and then we present the \bsc algorithm.

%

%For example, in several cases in the experiments, our basic (resp.\ fastest) approximation algorithm, took more than 5 hours  (resp.\  4 hours) to solve the problem in the default parameters setting, while more than 31 hours (resp.\ 8 hours)  in larger settings. 
%
%For example, in the experiments, our basic (resp.\ fastest) approximation algorithm, requires  2  to 6 hours  (resp.\ 40 min to 4.5 hours) to solve the problem in the default parameters setting, while 2.5 to 31 hours (resp.\ 1.5  to 8.5 hours)  in larger settings. 
%
%Hence, approximation algorithms are designed to solve the \ses problem.
%In what follows, we define the assignment score.

%The aforementioned imposed us to design approximation algorithms for solving the \ses problem.

%

 %
 % malon katlitera ..o kalitero approximation algorithmos mas ..pernei apo os xxx se default setting ..
%
%time bsc: in (YAHOO)  default settings 2 to 6 hours //  kai .max 2 - 31}
%horb 40min - 4.5  sto default kai /// 1.5 -8.5 sto max 

%\noindent
%Initially, we define the assignment score.

 \stitle{Assignment Score.}
Assume a schedule $\S$ and  an assignment $\alpha_{r}^t$, 
where $r$ is not previously assigned by $\S$ (i.e., $r \notin  \E(\S)$). 
As \textit{assignment score} (also referred  as \textit{score}) of an assignment $\alpha_{r}^t$, denoted as $\alpha_{r}^t.S$, 
we define the \textit{gain} in the expected attendance by including $\alpha_{r}^t$ in $\S$. 
The assignment score (based on Eq.~\ref{eq:user}) is defined as: 
 
\vspace{-0pt}
{ \small{
 \begin{equation}
  \label{eq:gain}
%\small  
 \alpha_{r}^t.S \,  \,  = \!  \! \underset{\substack{ \forall e_j \in  \\ \E_t(\S)\cup \{  r \}}}{ \sum} \!\!\!\!\!\!
  \omega'^{\, t}_{e_j}  
  \, - \,
  \underset{\substack{ \forall e_i \in  \\ \E_t(\S) }}{ \sum} \!
  \omega^{t}_{e_i}  
%  \textup{, where }
%  E'_t(\S) = \E_t(\S)\cup \{  e\}
  \end{equation}
   }}
% \vspace{-5pt}

\noindent 
Note that, the expected attendance   $\omega'$  
of each event after \mbox{assigning $r$,} differs from the  expected attendance  $\omega$ before the assignment. 
Also, based on Eq.~\ref{eq:user} \& \ref{eq:gain}, it is apparent that 
the score of an assignment referring to interval $t$ 
is determined based on all the  events assigned to $t$. 
Finally, given a set of assignments, the term \textit{top assignment} refers to the assignment with the largest score.

%
%\RED{Finally, given a set of assignments $\A$,  as $\tau(\A)$ we denote the assignment of $\A$ having the largest score, refereed as \textit{top assignment} of $\A$. } \BLUE{XRIAZETE????}

%\subsection{Greedy Algorithm  \textnormal{{(\bsc})}}
%\label{sec:bcs}

 \stitle{{Algorithm Outline}.}
Here we describe a simple   greedy   algorithm, referred as \textit{Greedy algorithm} (\bsc). 
The basic idea of \bsc is that the assignments between all pairs of event and interval are initially generated.
Then, in each step/iteration, the assignment with the largest score is selected. 
After selecting an assignment, a part of the potential assignment's scores have to updated.
Recall that the assignment's score is defined w.r.t.\ the events assigned in the assignment's interval (Eq.~\ref{eq:gain}).
Thus, when an assignment $\alpha_e^t$ is selected,  we have to recompute (update) 
 the scores of the assignments referring to $t$ interval. 
 
%\RED{ giati sigkrinomaste me ton basic? poli slow akoma k o basic ..opote ton exume san to kalitero result 
%} 

% \vspace{-6pt}
\stitle{Algorithm Description.}
Algorithm~\ref{algo:bsc} presents the pseudocode of  \bsc.
At the beginning the algorithm calculates the score,  
for all possible assignments (\textit{line}~3).
%between candidate events and intervals (\textit{line}~3).
The generated assignments are inserted into list $\L$  (\textit{line}~4). 
Then, in each step the assignment  $\alpha_{e_p}^{t_p}$ with the largest score is found and popped 
%(i.e., removed)  
(\textit{line}~6).
 If the popped assignment  is feasible and the event $e_p$ is not previous assigned 
 (i.e., assignment is valid), $\alpha_{e_p}^{t_p}$ is inserted into  schedule $\S$ (\textit{line}~8). 
%Otherwise, $\alpha_{e_p}^{t_p}$ is discarded.
Until a  valid assignment is selected, the assignment with the largest score  is popped and checked. 
After selecting  $\alpha_{e_p}^{t_p}$,
%the assignments in $\L$ that refer  to  interval  $t_p$ have to  be updated.
%The algorithm traverses $\L$  updating  the valid assignments referring to $t_p$ with the recomputed  scores%
the algorithm traverses $\L$,  updating  the appropriate assignments
%referring to $t_p$ with the recomputed  scores%
%
\alt{}{\footnote{\small We should note that,  the selection and the update of the assignments are performed (in parallel) though the same list traversal. 
However, for simplicity, in Algorithm~\ref{algo:bsc} are presented as different steps.}}
 (\textit{loop in line}~11). 
Finally, the algorithm terminates when $k$ assignments are selected.

 \stitle{Complexity Analysis.}
 %
%%%%%%%%%%%%%%%%%%%%%%%%%%%%
 %o BSC perform updates in  each iteration excepti the last 
 %so performs updates in E-i-1 events 0\leq i \leq k-2
%sto the num updates is 
%\sum_{i=0}^{k-2}   |\E| -  i-1   =
%$k|\E|+k/2 -k^2/2 -|\E|$
%%%%%%%%%%%%%%%%%%%%%%%%%%%%
%
%
\alt{}{In all the algorithms (omitted from the psedocodes) we assume that at the beginning  
the sum of the interest scores for the competing events is computed for each user and interval.  The sum is used in the assignment scores computations, in order to avoid 
accessing the competing events in each score computation.
The cost for the processing the competing events is $O(|\U||\C|)$.}%
Initially, the \bsc computes the  assignments for all event-interval pairs (\textit{loop in line}~2), 
which requires  $O(|\E||\T||\U|)$. 
Note that, each assignment score  (Eq.~\ref{eq:gain}) is computed in $O(|\U|)$.
In the next phase (\textit{loop in line}~5),  the \bsc performs $k$ iterations. 
In each iteration, the $\popAss$ operation (\textit{line}~6)  traverses the $\L$ list 
of size $|\T|(|\E|-i)$, where $0\leq  i \leq  k-1$. 
Thus, in sum, the cost for traversing $\L$ is 
%$O(\underset{{ i= 0 }}  {\overset{ k-1 } { \sum} }|\E||\T|-i)$.
$O(\sum_{i=0}^{k-1}|\T|(|\E|-i))$.
Additionally, in each iteration (except the last), $|\E|-(i+1)$ 
assignment updates are performed, in the worst case.
Hence,  the overall cost for updates is  
$O(|\U|\sum_{i=0}^{k-2}|\E|-i-1))$.
Therefore, the overall computation cost of \bsc in the worst case is 
%$O(|\U||\C|)+O(|\E||\T||\U|)+
%{O(\sum_{i=0}^{k-1}|\T|(|\E|-i))}+ 
%O(|\U|\sum_{i=0}^{k-2}|\E|-i-1)=$
$O(|\U||\C|+|\E||\T||\U|+k|\E||\T|+ k|\E||\U| - k^2|\T|-k^2|\U|)$.
Finally, the space complexity is $O(|\E||\T|)$.

 % \vspace*{-20pt} 
\begin{algorithm2e}[t]
\scriptsize
\setstretch{}

%\footnotesize
%\small
%\SetInd{0.5em}{1em}
%\SetVlineSkip {0.5mm} 
\caption{\bsc ($k$, $\T$, $\E$, $\C$, $\U$)}
\label{algo:bsc}
\KwIn{
$k$: number of scheduled events;  \: 
$\T$:  time intervals;  \linebreak
$\E$: candidate events;  \:  
$\C$: competing events;  \:  
$\U$: users;
}
\KwOut{$\S$: feasible schedule containing $k$ assignments}
 \KwVar{ $\L$:  assignment list}
\vspace{1mm}

$\S \gets \varnothing;$ \: $\L \gets \varnothing;$

 \ForEach(\Comment*[f]{\mycomment{{generate assignments}}}){$(e, t) \in  \E \times \T$}{
 	compute  $\alpha_e^t.S$
  \Comment*[r]{\mycomment{{{by Eq.\ \ref{eq:gain}}}}}
 	 insert $\alpha_e^t$ into $\L$
  \Comment*[r]{\mycomment{{{initialize assignment list}}}}

}

\While{$|\S|<k$}{

$ \alpha_{e_p}^{t_p}$ $\gets$ $\popAss(\L)$  \Comment*[r]{\mycomment{{{find \& remove the assignment with largest score}}}}

%	\If{$\alpha_{e_p}^{t_p}$ \text{is feasible} \Logand  $e_p \notin E(\S_k)$ }{
	\If{$\alpha_{e_p}^{t_p}$ \text{is \myvalid} }{
		insert $\alpha_{e_p}^{t_p}$ into $\S$
		  \Comment*[r]{\mycomment{{{insert the top and valid assignment into schedule}}}}

\If{$|\S|<k$}{

 \ForEach (\Comment*[f]{\mycomment{{update assignments}}})
 {$\alpha_{e}^{t} \in  \L$}{
	
%	\uIf{$\langle \alpha_{e}^{t} , \omega_{e}^{t} \rangle$  \text{is not feasible} \Logor $e \in E(\S_k)$ }{

	\uIf{$t = t_p$ \Logand $\alpha_{e}^{t}$  \text{is \myvalid} }{
compute new $\alpha_{e}^{t}.S;$  
%\: \: update $\alpha_{e_i}^{t_i}.S$ in $\L$
\Comment*[r]{\mycomment{{{by Eq.\ \ref{eq:gain}}}}}
 }
 
 	\lElseIf{$\alpha_{e}^{t}$  \text{is not \myvalid} }{remove $\alpha_{e}^{t}$ from $\L$;}
}
}
}
 }
 \Return $\S$\;
\end{algorithm2e}

%% file: eval.tex
%!TEX root = main.tex 

  \section{Experimental Analysis}
 \label{sec:eval}

  \vspace{-5pt}

  \subsection{Setup}
 \label{sec:setup}

 \stitle{Data.}
 In our experiments we use the largest \textit{Meetup dataset}  from \cite{Pham2015},  which contains data from {California}.
Adopting the same approach as in \cite{She2015,She2015a,She2016,Tong2015},  in order to define the interest of a user to an event,
we associate the events with the tags of the group who organize it.
Then, we compute the likeness value using Jaccard similarity over the user-event tags. 
After preprocessing, we have the  Meetup dataset  
containing   $42$,$444$ users and about $16$K events.
%, and the \meetupn, having  $32$,$444$ users and $10$K events. 
%Since the results of these datasets are similar, for brevity, we present only the results for   \meetupc, which is  the larger one.}
%Note that,  {similar results are also reported in} the second
%Meetup dataset from \cite{Pham2015} (referring  to New York).}
 % \vspace{-4pt}

 % 

{% 
\begin{figure*}[t]
\centering

\vspace*{-0.5cm}
\hspace*{-0.9cm}
  \subfloat[Utility vs.\ $k$ ]{\includegraphics[height=1.239in]{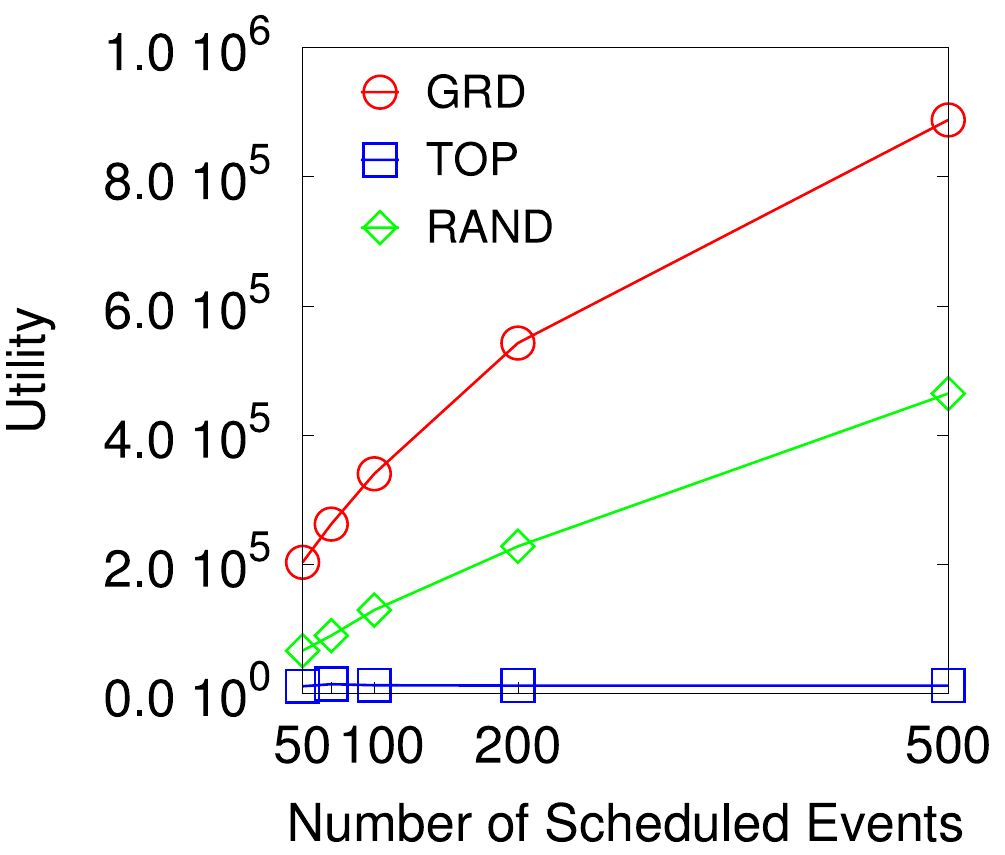}\label{fig:u_CA_k}}
  \hspace{0.5cm}
 \subfloat[Time vs.\ $k$ ]{\includegraphics[height=1.225in]{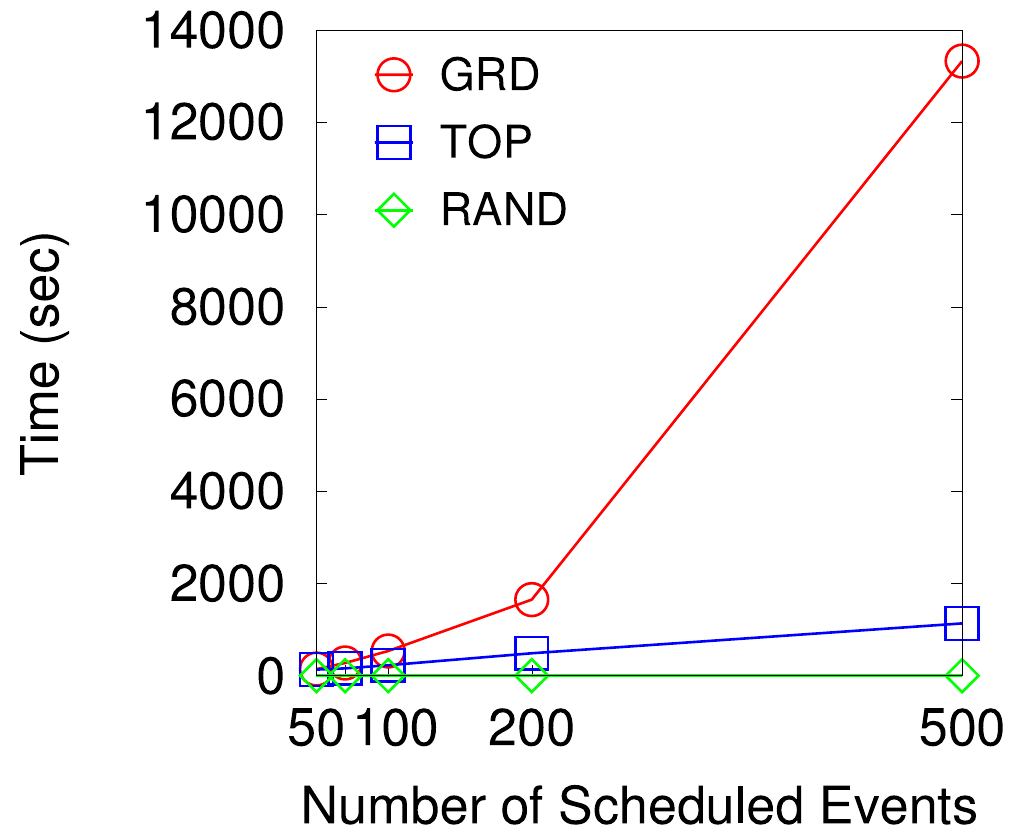}\label{fig:t_CA_k}}
  \hspace{0.5cm}
 \subfloat[Utility vs.\ $|\T|$ ]{\includegraphics[height=1.24in]{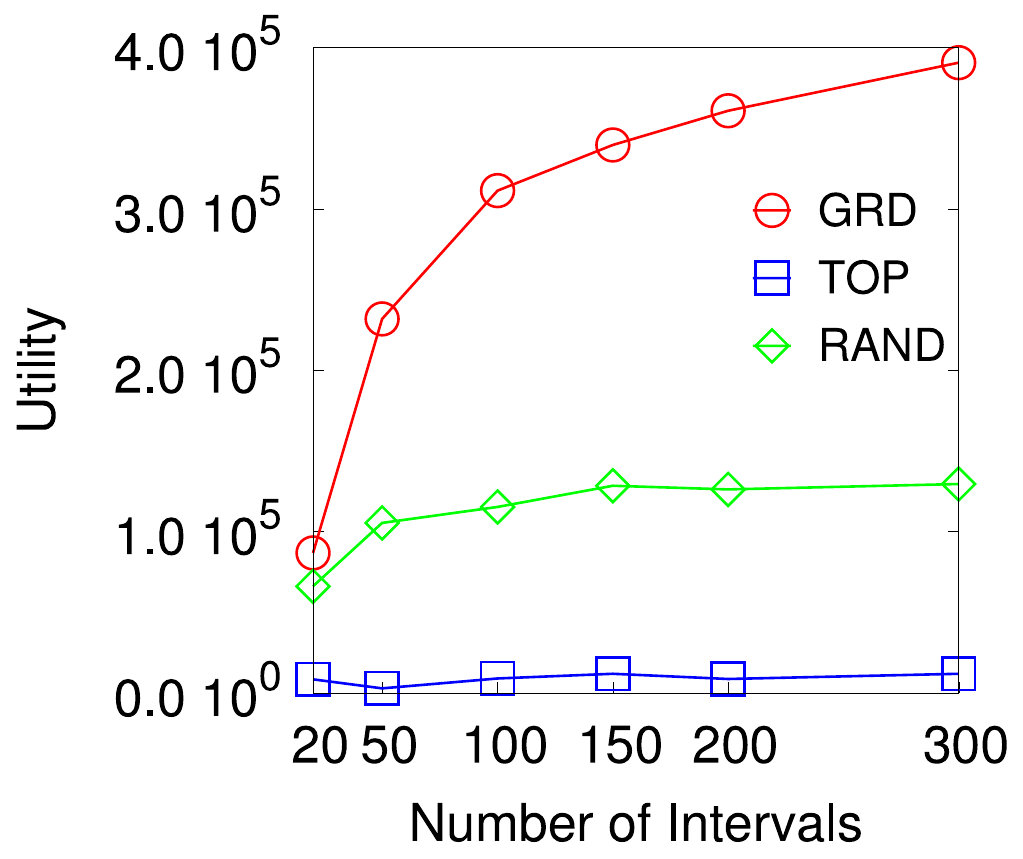}\label{fig:u_CA_i}}
 \hspace{0.5cm}
 \subfloat[Time vs.\ $|\T|$   ]{\includegraphics[height=1.237in]{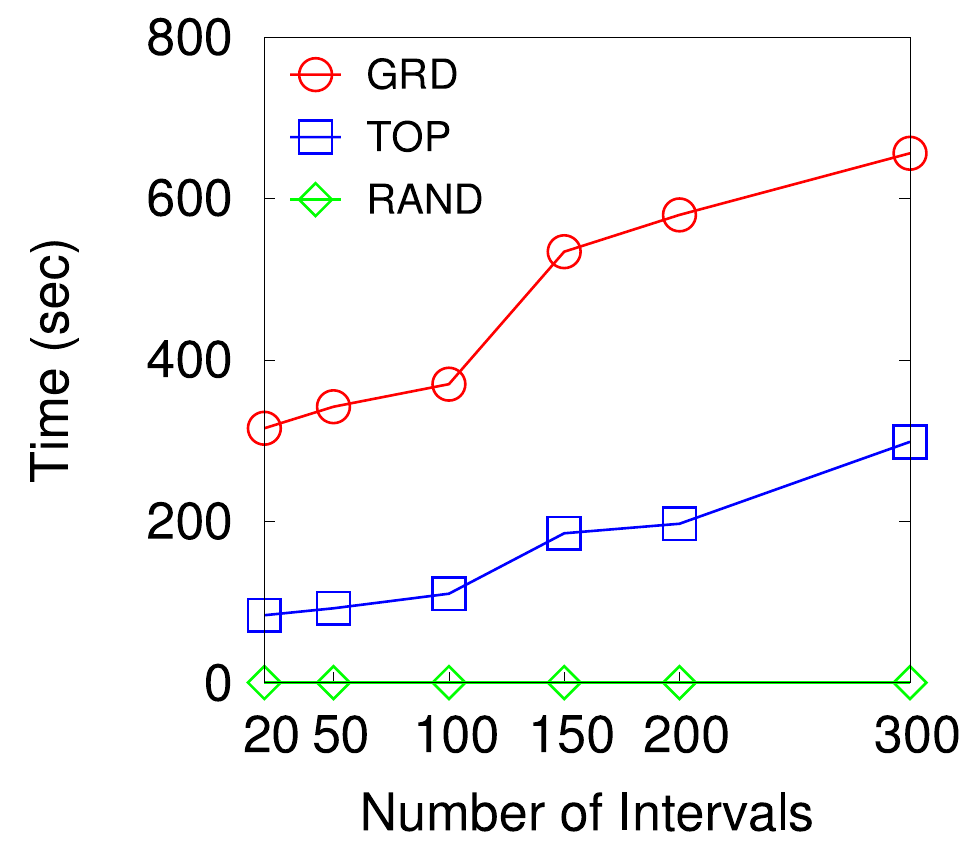}\label{fig:t_CA_i}}
 \hspace{0.0cm}
\caption{Varying the number of: (1) scheduled events  $k$; and   (2) time intervals $|\T|$}
\label{fig:vark}
%\vspace{-13pt}
\vspace{-9pt}
\end{figure*}
 } 

 \stitle{Parameters.}  
  Adopting the same setting as in the related works \cite{Li2014,She2015,She2015a,She2016,Tong2015}, 
we set the the default  and maximum value of the of \textit{scheduled events} $k$,  to $100$ and $500$, respectively.
Also, we vary the number of \textit{time intervals}  $|\T|$, from $\frac{k}{5}$ up to $3k$, with default value set to
$\frac{3k}{2}$.
Further, the number of \textit{candidate events} $|\E|$ is set to $2k$. 

 In order to select the values for  the number of competing events per interval, 
we analyze the two Meetup datasets from \cite{Pham2015}.   
From the analysis, we found that, on average, $8.1$ events are taking place during overlapping  intervals.
Therefore,  the number of \textit{competing events per interval} is 
selected by a uniform distribution having $8.1$ as mean value.

Regarding  the value for the number of {available events' locations},
we consider the percentage of pairs of  events that are spatio-temporally conflicting, as specified in  \cite{She2015}.
As a result, we set the number of \textit{available locations} to $25$.
 The \textit{social activity probability} $\sigma_u^t$ is defined using a Uniform distribution. 
 %Note that, the same results (not presented here) are also reported for Normal distribution.
 
The performance and the effectiveness of the examined methods are marginally affected 
by the available/required resources parameters (here as resources we consider organizer's staff).
Hence, we choose a reasonable number based on our scenario, setting the number of \textit{available resources}  to $20$. 
Also, the number of \textit{required resources} is selected by a uniform distribution defined over the interval $[1, \frac{20}{3}]$.

%
%\vspace{-4pt}
%\stitle{Metrics \& Implementation.}
%\alt{In each experiment, we measure: 
%$(1)$~the total utility score; 
%$(2)$~the execution time; and 
%$(3)$~the number of  computations for assignment scores  ($|\U|$ per assignment score).}
%%
%{In each experiment, we measure: 
%(1)~the total utility score, denoted as \textit{Utility}; 
%(2)~the execution time, denoted as \textit{Time}, and measured
%in secs; and 
%(3)~the number of  computations for assignment scores  ($|\U|$ per assignment score), denoted as \textit{Computations}.  
%}
%%
%The reported time values are the averages of 5 executions.
%All algorithms were written in C++ and the experiments were 
%performed on an 2.67GHz Intel Xeon  E5640 with 32GB of RAM. 
 
% The candidate
%products with the top-k values of the ranking functio

%
% 
\stitle{Methods.}
In our evaluation we study our method \bsc, as well as two baselines.
%\bsc (Sect.~\ref{sec:bcs}),  \bnd (Sect.~\ref{sec:bnd}), \hor (Sect.~\ref{sec:hor}) and \horb (Sect.~\ref{sec:horb}).
%Further, we also include two baselines.
The first baseline method, \stat, computes the assignment scores for all the events 
and selects the events with $top$-$k$ score values.
The second, denoted as \rand assigns events to intervals, randomly. 
%
 %Hence, since there are no score updates in \stat, \stat is always performing the minimum number of computations. 
%Beyond these methods,  we also include two baselines. 
%
Note that, since the objective,  the solution and  the setting of our problem 
are substantially different (see Sect.~\ref{sec:intro}) from the related works \cite{Li2014,She2015a,She2016,She2015,Tong2015,She2017,Huang08092016,ChengYCGW17}, 
the existing methods cannot be used to solve the \ses problem.
All algorithms were written in C++ and the experiments were 
performed on an 2.67GHz Intel Xeon  E5640 with 32GB of RAM.

%

%Further, we evaluate, \horh, a heap-based implementation of \hor, in which the $\L$ and $\M$ lists are replaced with max-heaps. For the heap structures we use the  standard implementation provided by the C++ STL library. 

 \subsection{Results}
\label{sec:results}

%\stitle{Effect of the Number of Scheduled Events.}
In the first experiment, we study the effect of varying \textit{the number of scheduled events} $k$. 
%Figure~\ref{fig:u_CA_k} presents the results, where the first row corresponds to 
%utility score, the second to number of  computations performed for assignment scores, and the third to execution time. 
%
%Figure~\ref{fig:u_CA_k} presents the results regarding utility score, and Figure~\ref{fig:t_CA_k} the execution time.
%
 In terms of  {\textit{utility}} (Fig.~\ref{fig:u_CA_k}),    we  observe that, in all cases, \bsc 
outperforms significantly both baselines. 
The difference  between   \rand and  \bsc increases as $k$ increases.
This is expected considering the fact that the larger the $k$, the larger the number of ``better'', compared to random, selected assignments. 
Finally,   \stat reports considerably low utility scores in all cases.
 
 The results regarding the \textit{execution time} are depicted in Figure~\ref{fig:t_CA_k}.
 Note that   the computations that are performed due to updates increase with $k$,  while the  number of {initially} computed scores is the same for all $k$.
Also,  \stat performs only the initial scores' computations (there are no score updates).  
That's why the difference between the \bsc and the \stat
increases with $k$.

 In the next experiment, we vary \textit{the number of time intervals} $|\T|$.
  Regarding  {\textit{utility}} (\mbox{Fig.~\ref{fig:u_CA_i}}),  we   observe that, as the number of intervals increases, the utility of \bsc and \stat methods increases too. 
This happens because the increase of available intervals  results to a smaller number of events
  assigned in the same interval, as well as to a larger number of candidate assignments. 
%The former results to  the assignment scores (in general) being larger in cases where fewer parallel events take place.
%The latter offers more options, which   possibly result to better assignments. 
%
 In terms of  \textit{execution time} (Fig.~\ref{fig:t_CA_i}), for the same reason  as in the first experiment, the difference between the \bsc and the \stat increases with $|\T|$.

%% file: concl.tex
%!TEX root = main.tex 

\section{Conclusions}
\label{sec:concl}

This paper introduced   the {\textit{Social Event Scheduling}} (\ses) problem. 
The goal of \ses is to maximize the overall events' attendance considering several events' and users' factors.
We showed that \ses is strongly NP-hard and we developed a greedy algorithm. 
 
\alt{}
{As future work, there are several variations that can be consider for the \ses problem.
in which new methods have to be designed.
For example, in a profit-based version, each event has a cost and a fee. 
In this scenario, an event cannot be organized if its expected profit 
(which can be computed w.r.t.\ expected attendance and fee) is not over a percentage of its cost.
Further,  users may be associated with a maximum number of attendances 
and/or a budget for a given time period.
In this case, a reasonable approach would consider for each user 
only her top attendances within the number attendances and budget limits.
}
\vspace{-5pt}